\documentclass[a4paper]{article}
\usepackage[margin=2cm]{geometry}
\usepackage[utf8]{inputenc}
\usepackage{bm}
\usepackage{graphicx}
\usepackage{gensymb}
\usepackage{url}
\usepackage{hyperref}
\usepackage{braket}
\usepackage{textcomp}
\usepackage{multirow}
\usepackage{subcaption}
\usepackage{color}
\usepackage{pdfpages}

\newcommand{\cm}[0]{cm$^{-1}$}
\newcommand{\gao}[0]{Ga$_2$O$_3$}
\newcommand{\bgao}[0]{$\beta$-Ga$_2$O$_3$}
\newcommand{\ggao}[0]{$\gamma$-Ga$_2$O$_3$}
\newcommand{\amode}[1]{A$_\mathrm{g}^{#1}$}
\newcommand{\bmode}[1]{B$_\mathrm{g}^{#1}$}
\newcommand{\fluence}[1]{$1\cdot 10^{#1}$ ions/cm$^2$}

\begin{document}

\title{Comprehensive structural and optical analysis of differently oriented Yb-implanted \bgao}
\date{}

\maketitle

\author{Joanna Matulewicz$^{1*}$, Renata Ratajczak$^{1}$, Mahwish Sarwar$^{2}$, Ewa Grzanka$^{3}$, Vitalii Ivanov$^{2}$, Damian Kalita$^{1}$, Cyprian Mieszczynski$^{1}$, Przemyslaw Jozwik$^{1}$, Slawomir Prucnal$^{4}$, Ulrich Kentsch$^{4}$, Rene Heller$^{4}$, Elzbieta Guziewicz$^{2}$}\\

${^1}$ National Centre for Nuclear Research, Andrzeja Sołtana 7, Otwock, Poland

${^2}$ Institute of Physics, Polish Academy of Sciences, Aleja Lotnikow 32/46, Warsaw, Poland

${^3}$ Institute of High Pressure Physics UNIPRESS, Polish Academy of Sciences, Sokolowska 29/37, Warsaw, Poland

${^4}$ Helmholtz-Zentrum Dresden-Rossendorf, Bautzner Landstrasse 400, Dresden, Germany


${^*}$ corresponding author: Joanna.Matulewicz@ncbj.gov.pl

\begin{abstract}

This study presents investigations of Yb-doped \bgao, an ultrawide bandgap semiconductor with potential use in future power and optoelectronic devices operating in high-radiation environments. The research has focused on the problem of structural damage caused by the implantation of Yb-ions into three differently oriented crystals and the optical response of created systems. The  (001), (010), and (-201)-oriented  \bgao\ crystals were implanted with three different fluences of 150 keV Yb ions and examined using a variety of experimental techniques: high-resolution X-ray diffraction (HRXRD), Rutherford backscattering spectrometry in channeling mode (RBS/c), Raman and photoluminescence (PL) spectroscopies, to provide comprehensive information about studied systems. Furthermore, the RBS/c studies were supported by Monte Carlo simulations. The results show distinctions between differently oriented crystals. In particular, (010)-oriented crystals are characterized by the lowest concentration of extended defects and the presence of compressive stress. In contrast, samples with the other two orientations exhibit tensile stress and significantly higher levels of extended defects. Interestingly, the PL spectra of (010)-oriented \bgao\ show the lowest emission from Yb\textsuperscript{3+} ions, suggesting that specific types of extended defects, whose formation is more favorable in the other two orientations than in (010), enhance Yb\textsuperscript{3+} luminescence instead of suppressing it.
\end{abstract}

\section{Introduction}

The development of materials has always played a key role in advancing technology and driving scientific progress. The discovery of new properties can open up new applications or improve the existing ones. One of particularly interesting materials nowadays, fulfilling the modern application market requirements, is a wide bandgap semiconductor - gallium oxide (\gao). It first captured the interest of researchers over 70 years ago, but in recent years it has seen a renewed interest and has been increasingly studied due to its remarkable properties. 

There are several polymorphs of gallium oxide:  $\alpha$, $\beta$, $\gamma$, $\delta$ , $\epsilon$ and $\kappa$  \cite{xu2022review}. Among them, only the monoclinic $\beta$ phase is thermodynamically stable \cite{higashiwaki2022beta}. Consequently, it is the commonly used polymorph, and it has been investigated in this work. With a bandgap of 4.8 eV \bgao\ belongs to the group of ultrawide bandgap semiconductors (UWBG). These materials are characterized by high critical breakdown field  \cite{xu2022review}.  These properties allow \bgao\ to handle high voltages. Moreover, it is chemically inert and can withstand high temperatures and other harsh conditions. 

Gallium oxide is optically transparent to visible and UV light, down to $\sim$250 nm \cite{pearton2018review, ping2021properties}, which can be particularly useful in optoelectronic devices. Furthermore, \bgao\ can be easily produced in bulk wafers of high quality, and the cost is lower than for common wide bandgap semiconductors such as SiC and GaN \cite{xu2022review,higashiwaki2022beta}. Because of its many advantages, the material has found numerous applications, including solar-blind UV photodetectors, efficient high-power switches \cite{baldini2018recent}, Schottky barrier diodes, metal-oxide-semiconductor field-effect transistors (MOSFET), and high electron mobility transistors (HEMT) \cite{xu2022review}. Due to its resistance gallium oxide is a promising material for harsh environments, such as space systems or nuclear power plants \cite{titov2022comparative}.

The luminescence range for undoped \bgao\ extends from ultraviolet to visible light \cite{ping2021properties}. Doping the material with rare earth ions (RE) such as ytterbium (Yb) extends this range into the infrared spectral range, which broadens the scope of applications \cite{sarwar2024defect}.  \bgao\ is considered a good host for rare earth ions, as it has been shown that the larger the band gap, the lower the thermal quenching effect \cite{favennec1989luminescence}.  Thus, Yb-doped \bgao, which is investigated in this work, has potential use in future optoelectronic devices such as high-power LEDs and high-end photodetectors operating even in high-irradiation environments.

Doping of \bgao\ with RE elements during the growth process frequently results in the formation of secondary phases or RE precipitation as a consequence of the low solubility limit of RE in this material and the high melting points of the dopants \cite{sadowski2007gaas,lide2004crc}.  Thus, one of the most common doping methods is ion implantation \cite{nikolskaya2021ion}, since the dopant concentration and depth profile can be precisely controlled by selecting an appropriate ion fluence and energy.  Moreover, implantation is a strongly non-equilibrium process allowing the introduction of any kind of atoms into solids with a concentration well above solid solubility limits. However, apart from many advantages, ion implantation can induce the formation of structural defects or phase transitions, which can alter the properties and the performance of the material. In the case of \bgao\ the effects of implantation have not been fully explored and are still under investigation.

\begin{figure}
    \centering
    \includegraphics[width=0.45\linewidth]{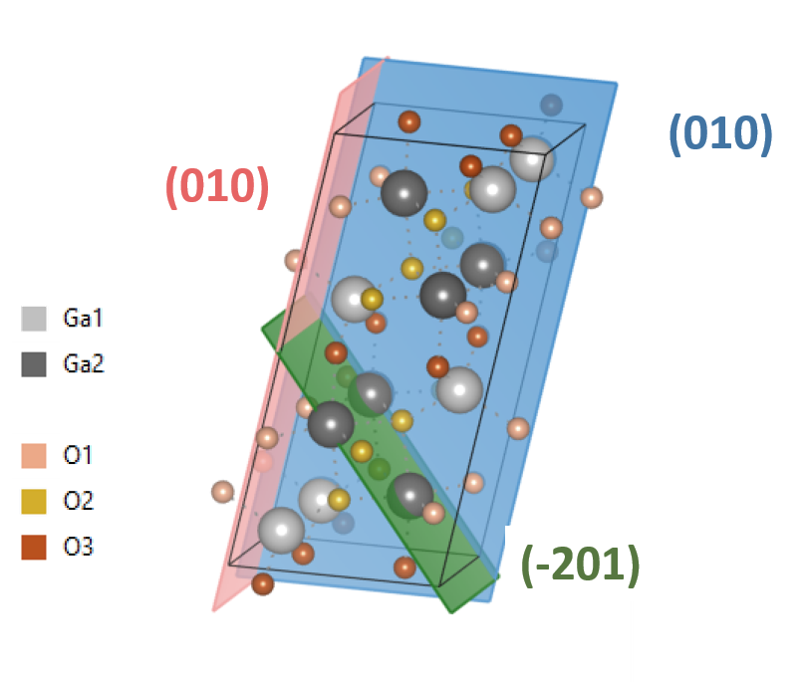}
    \caption{The unit cell of \bgao\ created by VESTA \protect\cite{momma2011vesta}. }
    \label{fig:cell}
\end{figure}

$\beta$-Ga$_2$O$_3$ belongs to the $C_{2h}^3$ ($C2/m$) space group \cite{janzen2022first}. The lattice parameters of monoclinic \bgao\ are reported as a = 12.23 Å, b = 3.04 Å, c = 5.80 Å and $\beta$ = 103.7$\degree$ \cite{kohn1957characterization}, and the unit cell is shown in Figure \ref{fig:cell}.  In the structure of \bgao\ two types of gallium and three types of oxygen atoms are distinguished, making up Ga$_\mathrm{I}$O$_4$ tetrahedrons and Ga$_\mathrm{II}$O$_6$ octahedrons.  Due to such a complicated structure and low symmetry, strong anisotropy is expected, which gave impetus to the studies of differently oriented crystals carried out in this work.

In this work, the radiation-induced structure changes in (001), (010), and (-201)-oriented \bgao\ implanted with different fluences of Yb ions were examined with a wide range of experimental methods: high-resolution X-ray diffraction (HRXRD), Rutherford backscattering spectrometry in channeling mode (RBS/c), Raman spectroscopy and photoluminescence spectroscopy (PL).  HRXRD scans illustrate crystal quality and its changes after implantation depending on crystal orientation. RBS/c combined with Monte Carlo simulations shows changes in defect concentrations and depth profiles. On the other hand, Raman spectroscopy provides information about the phonon energy distribution in the material, which reflects any changes in crystalline quality. Finally, the photoluminescence measurements are useful in analyzing the electronic and optical properties of the material and in determining its potential applications. These complementary methods deliver unique insights into the properties of Yb-doped \bgao\ systems, offering a comprehensive understanding of their structural, electronic, and optical behavior.

\section{Experimental Section}

\paragraph{Sample preparation}
Samples investigated in this work were (001), (010) and (-201)-oriented UID \bgao\ crystals from Novel Crystal Technology Inc. The crystals were n-type with electron concentration levels lower than $9\cdot10^{17}$ cm$^{-3}$. 
The samples measuring approximately 8x8 mm and 0.68 mm thick were obtained by dicing commercially available wafers.
A capping layer was deposited before the dicing, which was then etched using organic cleaning.

\paragraph{Implantation}
Crystals were implanted at room temperature with 150 keV Yb ions with fluences of $1\cdot10^{15}$, $1\cdot10^{14}$ and $1\cdot10^{13}$ ions/cm$^2$. To avoid channeled implantation the samples were tilted approximately 7$\degree$. The implantation processes were carried out at Helmholtz-Zentrum Dresden-Rossendorf (HZDR), Germany.
The selected implanted crystals were subsequently annealed in oxygen at 800\degree C for 10 minutes using Rapid Thermal Annealing (RTA) system  Accu Thermo AW-610 from Allwin21 Co. at the Institute of Physics, Polish Academy of Sciences (IP PAS), Warsaw, Poland.

\paragraph{XRD}
High Resolution X-Ray Diffraction (HRXRD) measurements were
performed at UNIPRESS Institute of High Pressure Physics, Polish Academy of Sciences, Warsaw, Poland, with Empyrean X-ray diffractometer operating at Cu K$\alpha$1 wavelength (40kV, 30mA), equipped with a hybrid 2-bounce monochromator and a threefold Ge(220) analyzer. For each sample 2$\theta$/$\omega$ scan of symmetrical reflection were done to measure the implantation effect. 

\paragraph{RBS/c}
The RBS/c measurements were performed at HZDR, Dresden, Germany. For each sample two different modes of measurements were applied: random and aligned , i.e. with the crystal oriented randomly and aligned along the surface orientation. A beam of 1.7 MeV He$^+$ ions was delivered with a Van de Graaff accelerator. A silicon detector with a scattering angle of 170$\degree$, depth resolution of about 5 nm and energy resolution of less than 20 keV was used.

\paragraph{Raman spectroscopy}
The Raman spectroscopy measurements of \bgao\ were performed at HZDR, Dresden, Germany. The phonon spectra have been obtained in backscattering geometry in the range of 50 to 1000 \cm\ using a 532 nm Nd:YAG laser with a liquid nitrogen-cooled charge-coupled device camera. Eight selected samples were examined by Raman spectroscopy.

\paragraph{PL spectroscopy}
Photoluminescence (PL) spectroscopy measurements of Yb implanted and annealed \bgao\ were carried out at IP PAS at T=300 K under near-band edge excitation at 5.82 eV with an Edinburgh Instrument FLS 1000 fluorescence spectrometer and a CCD3-x camera to collect the spectra at the wide range from 1.17 to 4.97 eV and an InGaAs camera (DU49\_X) to collect spectra only in the near-infrared (NIR) spectral range. Yb-related inner shell emissions were detected in single photon counting mode in the range of 1.20-1.28 eV.

\section{Results and discussion}

\subsection{HRXRD}

HRXRD scans provide detailed information about the crystalline quality of the material and its changes after ion implantation. The asymmetry of the main peaks and the appearance of additional ones may indicate the development of strain, amorphization, or formation of other crystal phases in the studied material. 

\begin{figure}[h!]
    \centering
    \includegraphics[width=0.85\linewidth]{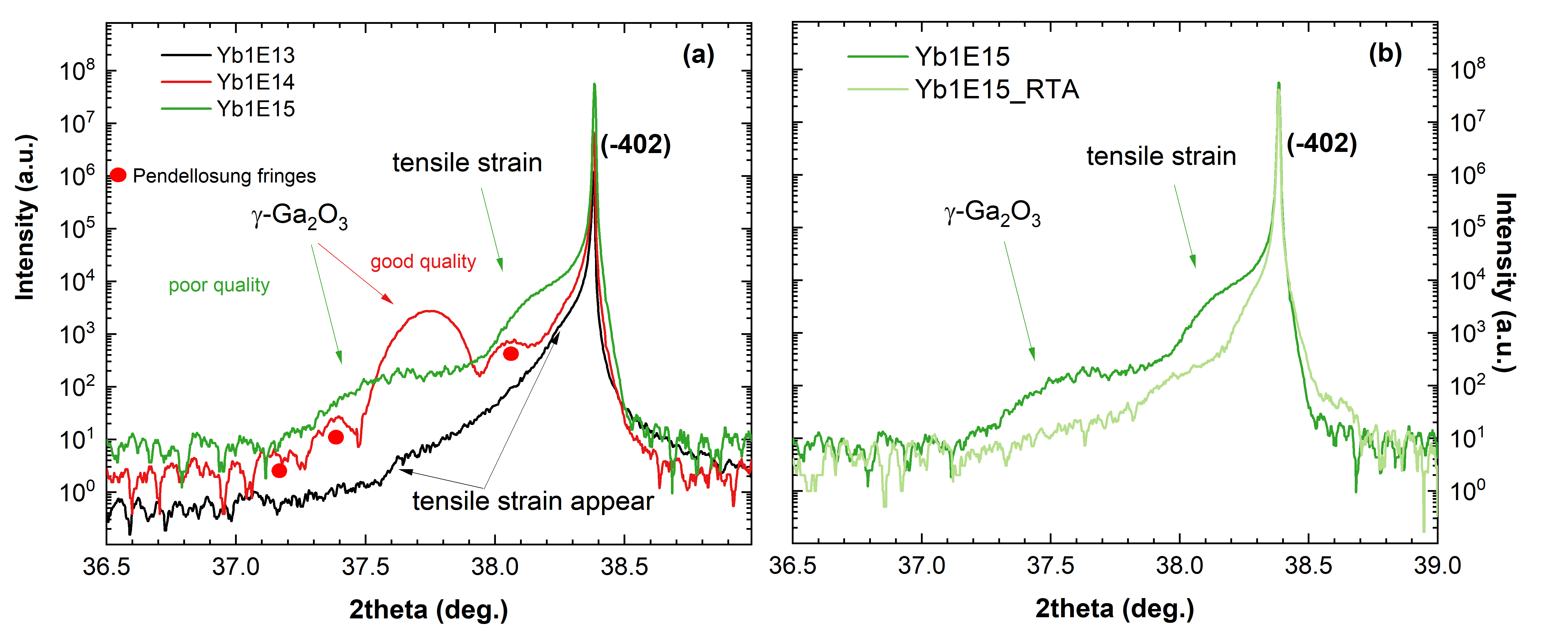}
    \caption{HRXRD 2$\theta$ scans of (-201)-oriented \bgao\ crystal (a) implanted with Yb ions with different fluences and (b) implanted with Yb with fluence of \fluence{15} and annealed in oxygen at 800\degree C for 10 minutes. }
    \label{fig:xrd_201}
\end{figure}

The HRXRD results for (-201)-oriented \bgao\ crystal implanted with three fluences of $1\cdot10^{13}$,  $1\cdot10^{14}$ and \fluence{15} of Yb are presented in Figure \ref{fig:xrd_201}a. The sharp signal coming from the symmetric (-402) reflection of \bgao\ crystals implanted with the lowest fluence has a full width at half maximum (FWHM) calculated from 2$\theta$ scans equal to 0.008$\degree$. This is consistent with the results calculated for the virgin \bgao\ crystal from the rocking curve, presented in our previous work \cite{sarwar2024defect}, and indicates that the quality of the initial crystals is very high. 
However, the asymmetry of the (-402) reflection towards increased lattice parameters due to tensile strain is also visible, which indicates the beginning of the structural changes for a sample with the lowest implantation fluence. Distinct evolution of the central (-402) peak at 38.38$\degree$ occurs with increasing implantation fluence and shoulder peaks at 38.3$\degree$ and 37.7$\degree$ are clearly visible. The first shoulder peak is related to implantation-induced strain \cite{sarwar2024defect, kjeldby2022radiation} while the second one appears due to radiation-induced phase transformation from $\beta$ to $\gamma$ phase \cite{sarwar2024defect,azarov2024optical,yoo2022atomic}.
In the crystals implanted with the middle fluence, i.e. \fluence{14}, the radiation-induced phase transformation from $\beta$- to \ggao\ phase is evident. As can be seen in Figure SM1 in the Supplementary Materials, the $\gamma$ phase was developed in the entire implanted zone. In addition, the presence of oscillations in the diffraction pattern, coming from the \ggao\ layer, allows us to determine its average thickness to be approximately 40 nm.  
In the crystals implanted with the highest fluence, i.e. \fluence{15} two additional shoulder peaks remain visible. However, comparing this diffraction pattern with the one obtained for the middle fluence used (\fluence{14}) reveals that the peak coming from \ggao\ phase becomes wider and its intensity decreases. The peaks shift their positions toward the lower 2$\theta$ angles (higher lattice parameters). This indicates a deterioration of the structural quality, including the beginning of the amorphization process as the implantation fluence increases up to \fluence{15}. 
These observations were confirmed by Nano-Beam electron Diffraction (NBD) patterns taken from the selected layers visualized by HRTEM image (see Figure SM2 in Supplementary Materials, with a detailed description in the work of Ratajczak et al. \cite{ratajczak2024anisotropy}). It is worth noting that after annealing, the radiation-induced $\gamma$ phase of gallium oxide disappears and the strain is reduced, as can be seen in Figure \ref{fig:xrd_201}b and Figure SM3. 

\begin{figure}[h]
    \centering
    \includegraphics[width=1\linewidth]{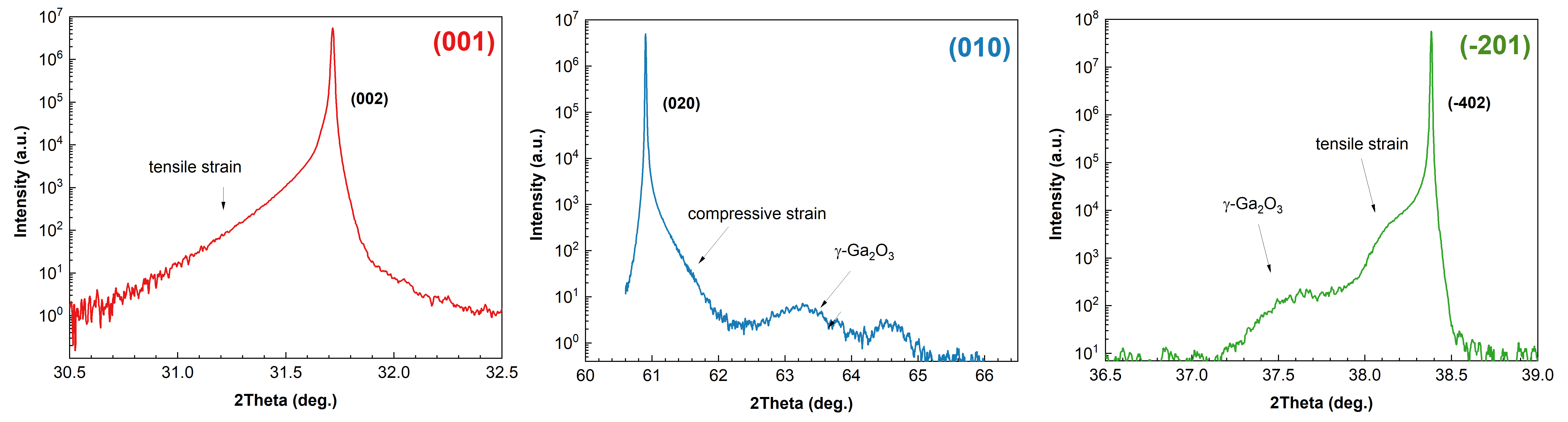}
    \caption{HRXRD 2$\theta$ scans of (-201)-oriented \bgao\ crystal implanted with Yb ions with a fluence of \fluence{15} } 
    \label{fig:xrd}
\end{figure}

HRXRD scans presented in Figure \ref{fig:xrd} show how the crystalline structure of differently oriented \bgao\ changes after implantation with the highest Yb fluence. As can be seen, in the case of both (001) and (-201)-oriented crystals the evolution of the peak suggests implantation-induced tensile strain. In contrast,  for (010)-oriented crystals compressive strain is observed \cite{kjeldby2022radiation}. In addition, in the diffraction patterns for the (010) and (-201)-oriented crystals additional peaks associated with the formation of \ggao\ can be observed \cite{sarwar2024defect, azarov2024optical}. According to the work of Gottschalch et al. \cite{gottschalch2019heteroepitaxial}, in the angular region of the diffractogram collected for the (001)-oriented crystal there are no peaks originating from \ggao.

\subsection{RBS/c}

Rutherford backscattering spectrometry measures the energy and number of analyzing ions scattered back from the atoms of the investigated material. This method allows for depth analysis of the stoichiometry and the impurities near the surface (up to about 1 $\mu$m).  
In the case of Yb-implanted \bgao, all three elements present differ greatly in atomic mass. Hence, the signals coming from ions scattered on Ga, O, and Yb atoms can be easily distinguished and analyzed separately.

For RBS measurements performed in the channeling mode (RBS/c), when the beam is aligned perpendicular to the crystal surface, the backscattering yield is much lower than in the random mode. It is due to the channeling effect \cite{chu1978backscattering}. For both the undoped and implanted crystals a surface peak appears, which corresponds to ions scattered from surface atoms. Because of the defects induced by the implantation process, the backscattering yield for implanted crystals is higher than for undoped ones. It is worth mentioning that different types of defects affect the spectra in different ways.  Simple defects induce the direct backscattering of analyzing ions and cause the growth of the characteristic damage peak, while extended defects result in an increased backscattering yield behind the damage peak, due to the mechanism called dechanneling, which results in delayed scattering \cite{caccador2023extracting}. This term is also commonly used to refer to the yield behind the damage peak itself. 

\begin{figure}[h!]
    \centering
    \includegraphics[width=0.65\linewidth]{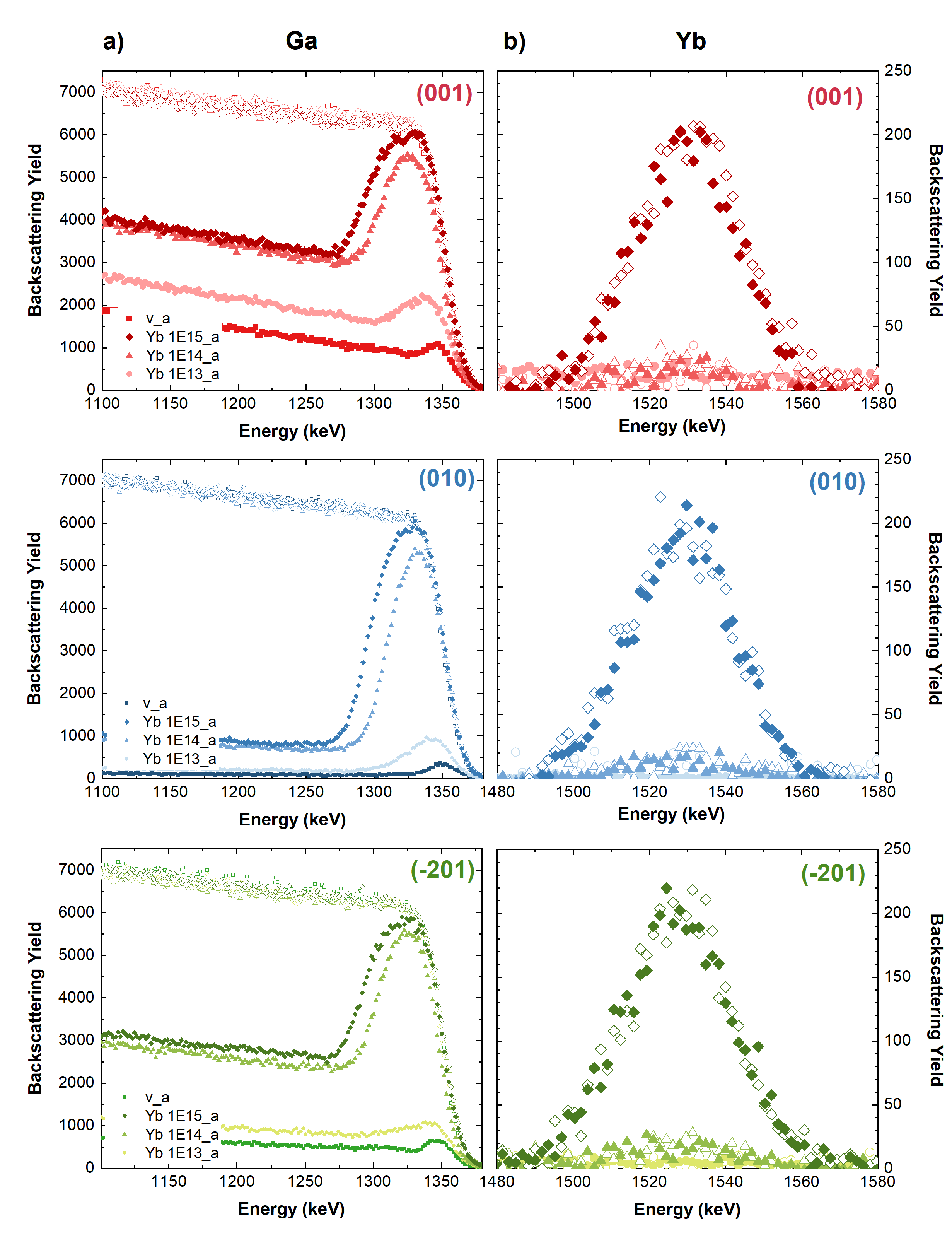}
    \caption{RBS random (open symbols) and aligned (solid symbols) spectra for differently oriented \bgao\ crystals implanted with different fluences of Yb ions. The spectra on the left (a) show the signal from gallium atoms and the spectra on the right (b) - from ytterbium atoms.}
    \label{fig:rbs}
\end{figure}

\begin{figure}[h!]
    \centering
    \includegraphics[width=0.25\linewidth]{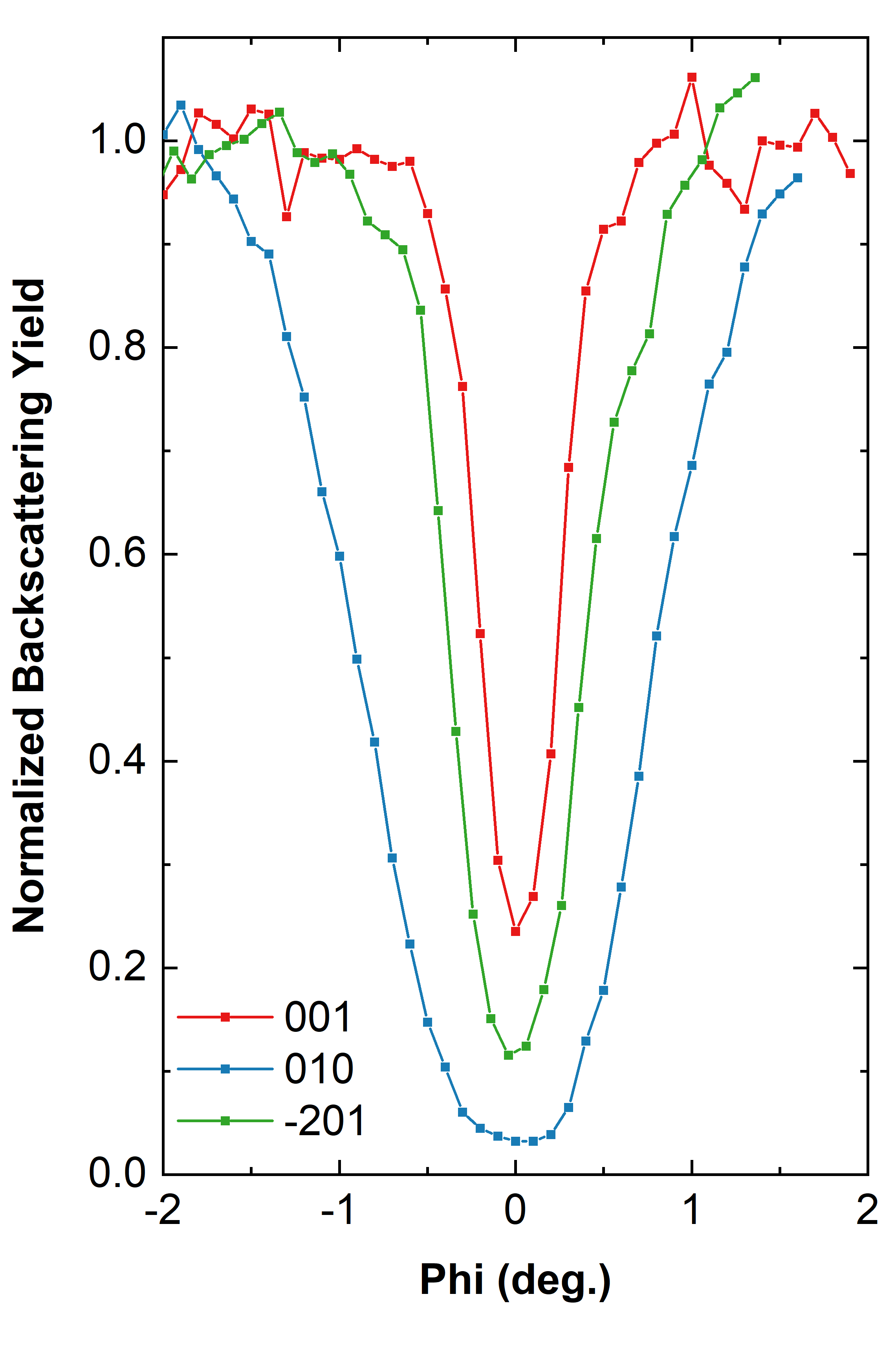}
    \caption{The angular scans obtained by RBS/c analysis of virgin samples of (001), (010), and (-201)-oriented \bgao\ crystals.}
    \label{fig:rbsscan}
\end{figure}

The RBS/c spectra, presented in Figure \ref{fig:rbs}, show the number of ions scattered on gallium atoms (a) and ytterbium atoms (b) in differently oriented \bgao. The signal from oxygen that appears at the lowest energy range (below the energy of 616 keV) is barely visible due to the low cross-section for scattering on light elements and was not analyzed. The spectra in the energy range corresponding to Ga provide information about the defect distribution and those obtained for Yb show the dopant lattice site location and its depth profile. As expected, the characteristic damage peak in Figure \ref{fig:rbs} increases with fluence. The defect accumulation process appears to be very similar for all tested crystals. However, significant differences in the dechanneling levels can be observed. This is partly due to the different sensitivity of the RBS/c method on different surface orientations of the crystals because the crystal channels do not look the same (see Figure \Ref{fig:rbsscan} and the work of Anam et al. \cite{anam2021structural}). Therefore, significant differences can be seen even in the aligned spectra obtained for undoped crystals. Nevertheless, the performed simulations described below show that it does not fully explain the differences observed for Yb-implanted differently oriented crystals.

\begin{figure}[h!]
    \centering
    \includegraphics[width=0.5\linewidth]{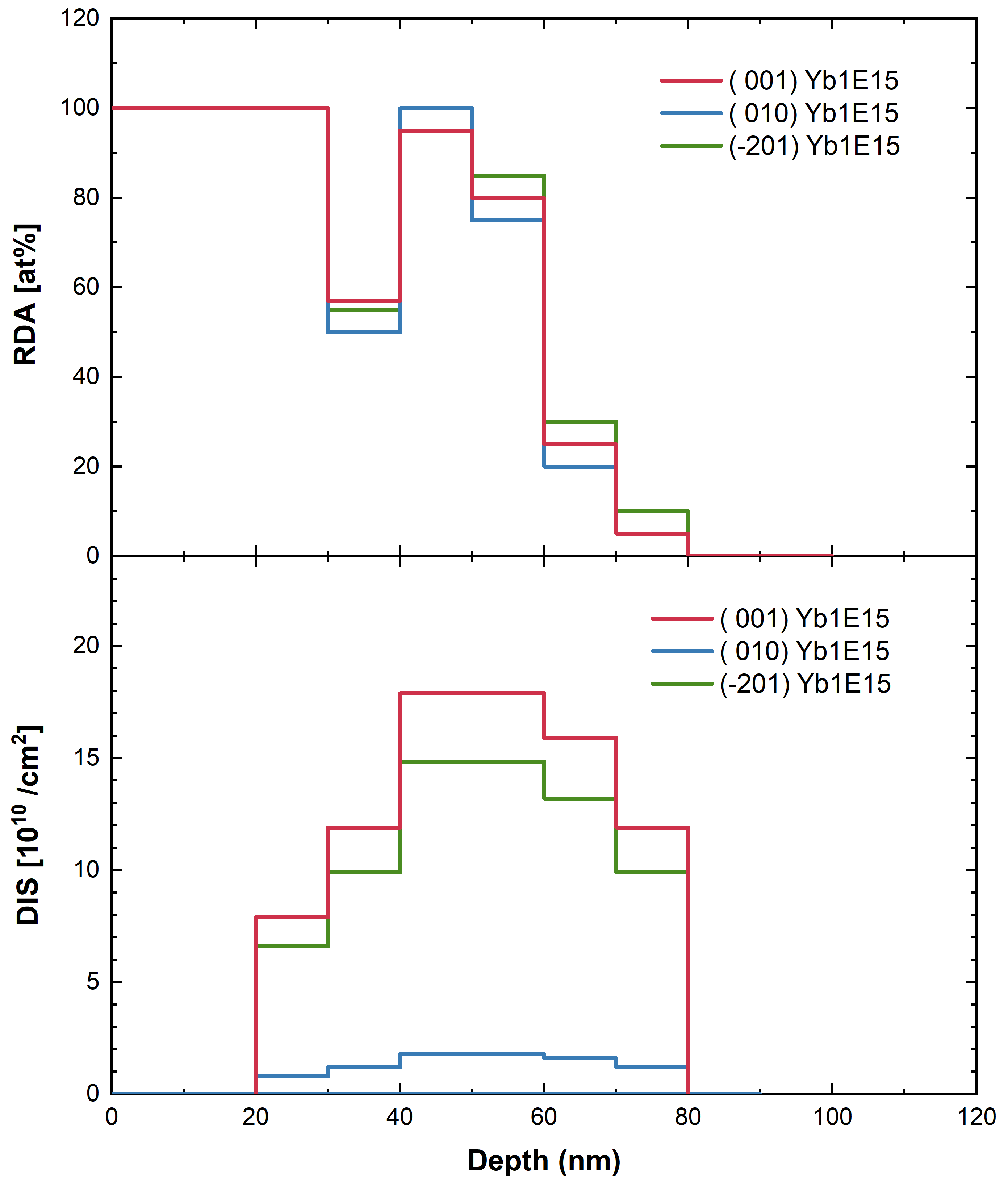}
    \caption{Concentrations of simple defects (RDA) and extended defects (DIS) obtained by McChasy for the differently oriented \bgao implanted with Yb to the fluence of \fluence{15} }
    \label{fig:rda_dis}
\end{figure}

In order to better understand the RBS/c spectra, McChasy simulations were performed to obtain information about defect types and their distributions. McChasy is a Monte Carlo simulation tool, which reproduces the path of a projectile in the given material \cite{jozwik2023monte,jozwik2020advanced}. The distributions of two distinct types of defects can be obtained by McChasy: simple and extended ones. In the program, the simple defects, which act as channel-blocking defects and lead to direct backscattering of the analyzed ions,  are modeled as randomly displaced atoms (RDA). In turn, extended defects (named DIS) refer to those that cause the bending of crystal channels like edge dislocations, clusters, or grain boundaries, causing the local distortion effect of the crystal lattice. It should be pointed out that the reorganization of atoms due to the changes in phase or orientation compared to the bulk crystal, reported in the work of Azarov et al. \cite{azarov2024optical} is also viewed as a defect according to McChasy. The same contribution to the RBS-aligned spectrum derived by different defects or other structural changes makes the full identification impossible. For this reason, a variety of techniques were used, which have their own limitations but are complementary to each other.

The results of McChasy simulations performed for the differently oriented \bgao\ crystals implanted with the highest fluence, i.e  \fluence{15}, are presented in Figure \ref{fig:rda_dis}. As can be observed, the distribution of RDA-type defects is relatively consistent for all considered crystal orientations of \bgao, whereas the concentration of bending channels defects is notably lower for (010)-oriented crystal compared to (001) and (-201). This difference is the main reason for the lower dechanneling level observed in the aligned spectra behind the damage peak for the implanted crystal. It is also worth mentioning that the distribution of RDA becomes bimodal for fluences exceeding \fluence{14} due to the formation of two phases in the implanted crystals: amorphization and $\gamma$ phases  \cite{ratajczak2024anisotropy}(see Figure SM2).

Additionally, as can be seen in Figure \ref{fig:rbs}b, the Yb behavior is similar for all surface orientations. Its signal in the aligned spectra is at the same level as in the random spectra for the visible fluences. However, due to the $\beta-\gamma$ phase change occurring in the modified layer after implantation, no conclusions can be drawn regarding the position of the Yb ions sites in the \ggao\ crystal lattice.

\subsection{Raman spectroscopy}

Raman spectroscopy is a technique based on inelastic scattering of photons. The energy of the incident photon may decrease or increase as a result of excitation or deexcitation of vibrations. Since the energies of incident and scattered photons are close and the difference between them is measured, the incident radiation must be monochromatic. Therefore, a laser is used for measurements, and the Raman shift expressed in \cm, i.e. in the unit of the wave vector, is marked on the spectrum axis. 

Characteristic Raman spectra contain peaks, which correspond to different vibration modes of microstructures in the studied material - phonon modes. The positions of peaks may differ from ideal under the influence of such factors as temperature, stress, or when the material contains structural defects and impurities. The presence of defects can also affect the Raman lineshape \cite{bergman1996raman}. 

\begin{figure}[h!]
    \centering
    \includegraphics[width=0.9\linewidth]{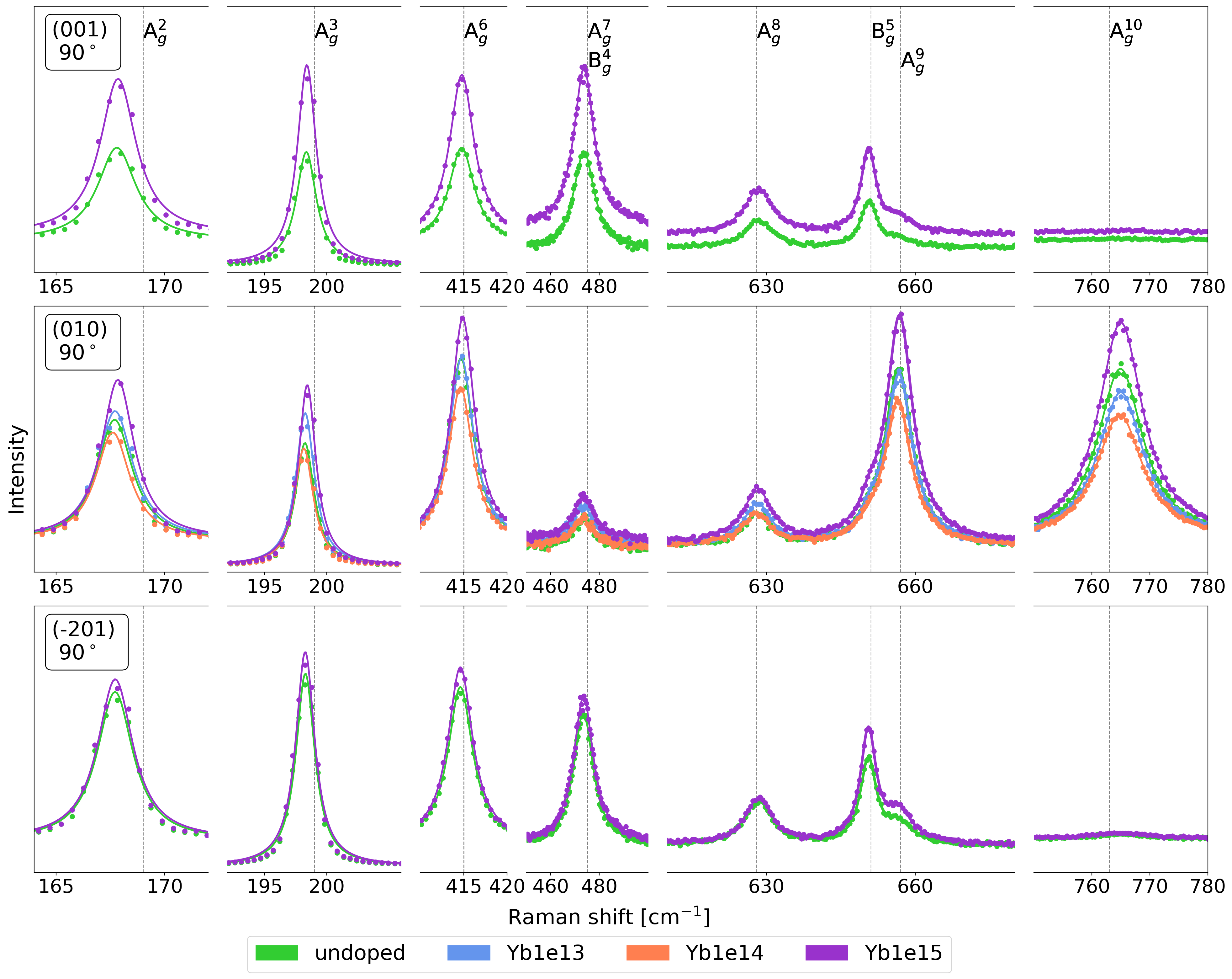}
    \caption{Raman spectra of undoped and Yb-implanted \bgao. The scale is different for each part of the spectrum, but it is the same for different surface orientations. The experimental data is marked with dots and the solid line indicates the fit to the Lorentz distribution. Phonon modes with values from the work of Dohy et al. \protect\cite{dohy1982raman} are marked with dashed lines. }
    \label{fig:raman}
\end{figure}

The primitive unit cell of \bgao\ has four gallium and six oxygen atoms, resulting in 30 phonon modes in the first Brillouin zone \cite{janzen2022first,kranert2016raman}. The irreducible representation in $\Gamma$ point is
\[
\Gamma = 10 A_g + 5 B_g + 5 A_u + 10 B_u,
\]
where modes with index $g$ are Raman-active and modes with index $u$ are IR-active. Therefore, ten $A_g$ and five $B_g$  active phonon modes are expected in the Raman spectroscopy measurements for \bgao\ crystals. These phonon modes are listed in Table \ref{tab:peaks}. The collected Raman spectra, presented in Figure \ref{fig:raman} show signals from almost all of these phonon modes of \bgao.

\begin{table}[h!]
\centering
\caption{Raman shifts in \cm\ of phonon modes from literature \protect\cite{dohy1982raman} and from this work in the undoped (010)-oriented crystal}
\begin{tabular}{p{1cm}c|c|c|c|c|c|c|c|c|c|c|c|c|c|c|c|}
\hline
\multicolumn{2}{|c|}{Phonon mode}                         & $A_g^1$ & $B_g^1$ & $B_g^2$ & $A_g^2$ & $A_g^3$ & $A_g^4$ & $A_g^5$ & $B_g^3$ & $A_g^6$ & $A_g^7$ & $B_g^4$ & $A_g^8$ & $B_g^5$ & $A_g^9$ & $A_g^{10}$ \\ \hline
\multicolumn{1}{|c|}{\multirow{2}{*}{\begin{tabular}{@{}c@{}}Raman \\ shift\end{tabular}}} & Literature  & 111 & 114 & 147 & 169 & 199 & 318 & 346 & 353 & 415 & 475 & 475 & 628 & 651 & 657 & 763 \\ \cline{2-17} 
\multicolumn{1}{|c|}{}                    & This work & 108 & 112 & 142 & 167 & 198 & 318 & 344 & - & 414 & 473 & 473 & 628 & 650 & 656 & 764 \\ \hline
\end{tabular}
    \label{tab:peaks}
    
\end{table}

The peaks on the Raman spectra were fitted to the Lorentz distribution with a flat background. The calculated positions for undoped (010)-oriented crystal are also listed in Table \ref{tab:peaks}. The Raman peak positions, obtained in this work are similar to those reported in the literature \cite{janzen2022first,kranert2016raman,dohy1982raman,onuma2014polarized}. Moreover, no additional peaks were observed, which means that only \bgao\ structure was noted.

\begin{figure}
    \centering
    \includegraphics[width=0.75\linewidth]{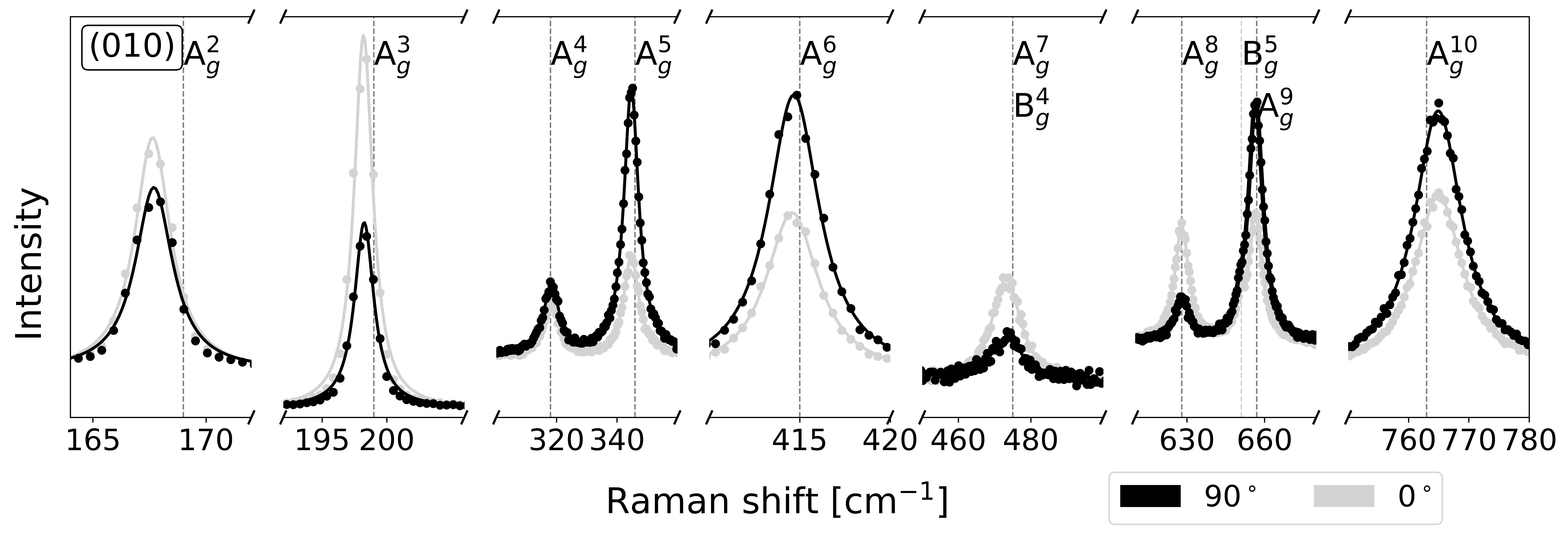}
    \caption{Raman spectra of (010) undoped \bgao\ crystal rotated 90\degree.}
    \label{fig:raman_angle}
\end{figure}

The collected Raman spectra showed variation in the intensity of certain modes depending on the sample rotation in the horizontal plane.
The samples were measured at three rotation positions 0, 45, and 90\degree. In the case of the 45\degree  angle, no regular changes were observed so no further analysis was performed for such spectra. For 0$\degree$ and 90$\degree$ angles the spectra differed particularly in the ratio of the intensity of \amode{8} and \amode{9}/\bmode{5} modes and \amode{4} and \amode{5}/\bmode{3}, which can be seen in Figure \ref{fig:raman_angle}. These differences allowed us to assign the angles to measurements in order to compare results for the same polarization.  The spectra measured for samples rotated 90$\degree$ were chosen for further analysis.

\begin{figure}[h!]
    \centering
    \includegraphics[width=0.35\linewidth]{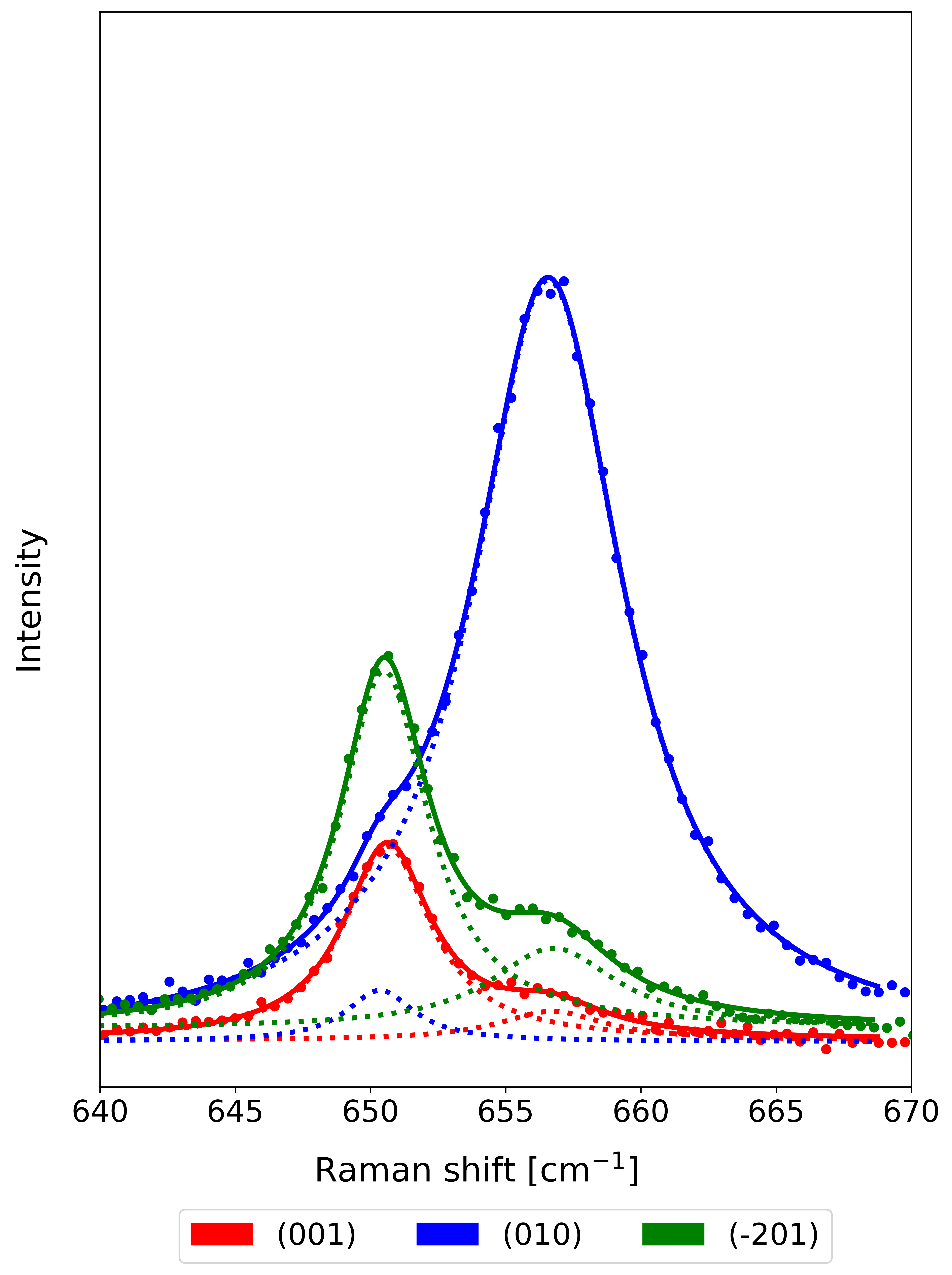}
    \caption{\amode{9} (657 \cm) and \bmode{5} (651 \cm) modes in a function of surface orientations of the crystal fitted as a sum of two Lorentz distributions. Dotted lines correspond to the two components of the sum. }
    \label{fig:lorentz_sum}
\end{figure}

\begin{table}[h!]
\centering
\caption{Amplitudes and FWHM of \amode{9} and \bmode{5} modes for undoped crystals.}
\begin{tabular}{|c|cc|cc|c|}
\hline
\multirow{2}{*}{Surface orientation} & \multicolumn{2}{c|}{\amode{9}}        & \multicolumn{2}{c|}{\bmode{5}}        & \multirow{2}{*}{Ratio of amplitudes \amode{9}/\bmode{5}} \\ \cline{2-5}
                                     & \multicolumn{1}{c|}{Amplitude} & FWHM [\cm] & \multicolumn{1}{c|}{Amplitude} & FWHM [\cm] &                                                          \\ \hline
(001)                                & \multicolumn{1}{c|}{0.46}      & 4.8  & \multicolumn{1}{c|}{3.07}      & 4.0  & 0.15                                                     \\ \hline
(010)                                & \multicolumn{1}{c|}{12.04}     & 6.8  & \multicolumn{1}{c|}{0.81}      & 3.1  & 15                                                       \\ \hline
(-201)                               & \multicolumn{1}{c|}{1.26}      & 5.7  & \multicolumn{1}{c|}{5.65}      & 4.0  & 0.22                                                     \\ \hline
\end{tabular}
\label{tab:lorentz_sum}
\end{table}

Differently oriented \bgao\ crystals show the presence of different phonon modes and peak intensities, as shown in Figure \ref{fig:raman}. In particular,  in the case of (001) and (-201)-oriented \bgao\ the \bmode{5} mode is dominant over \amode{9} mode. In contrast, for (010)-oriented crystal the \amode{10} mode is present and the \amode{9} mode has a much higher intensity than \bmode{5} (see Figure \ref{fig:lorentz_sum}, Table \ref{tab:lorentz_sum}). As can be found in the literature \cite{zhang2021raman} \amode{10} corresponds to the bending vibration of the Ga$_\mathrm{I}$(O$_\mathrm{II}$)$_2$ bonds and to the symmetrical stretching vibration of the Ga$_\mathrm{I}$(O$_\mathrm{I}$O$_\mathrm{III}$)$_2$ bonds of the Ga$_\mathrm{I}$O$_4$ tetrahedrons. \bmode{5} is assigned to the asymmetrical stretching vibration of the Ga$_\mathrm{I}$(O$_\mathrm{II}$)$_2$ bonds of the same tetrahedral unit.  Furthermore, the signal from \amode{7} and \bmode{4} has a considerably lower intensity for (010)-oriented crystal. \amode{7} is associated with stretching and bending of Ga$_\mathrm{I}$O$_4$ as well \cite{zhang2021raman}. 
Even though the collected Raman spectra do not show any characteristic response to strain, such as shifts of the modes to the right or left depending on the type of strain \cite{myers1983effects}, the differences in the intensities of these peaks indicate specific properties of the studied material. 

Moreover, one can observe the differences in the intensity of the modes for different fluences. Although the differences are not significant, a decrease in the intensity with the increasing fluence is noticeable, due to the damage \cite{myers1983effects}. However, for the highest Yb fluence of \fluence{15} the amplitude of the peaks is higher than for the virgin one, in the case of all tested crystal orientations, and the FWHM is the lowest for most of the phonon modes. It may be hypothesized that this phenomenon is associated with the plastic deformation of the crystal with the clearly observed threshold in the vicinity of this specific fluence \cite{sarwar2024defect}.

\subsection{PL spectroscopy}

PL spectroscopy is a powerful tool for studying the electronic structure, local symmetry, excitation, and relaxation dynamics of emission centers associated with local defects and impurities in solids \cite{shinde2012phosphate}. This technique was applied here to study the optical properties of Yb\textsuperscript{3+} ions implanted in \bgao\ crystals with different crystallographic orientations.

\begin{figure}[h!]
    \centering
    \includegraphics[width=0.8\linewidth]{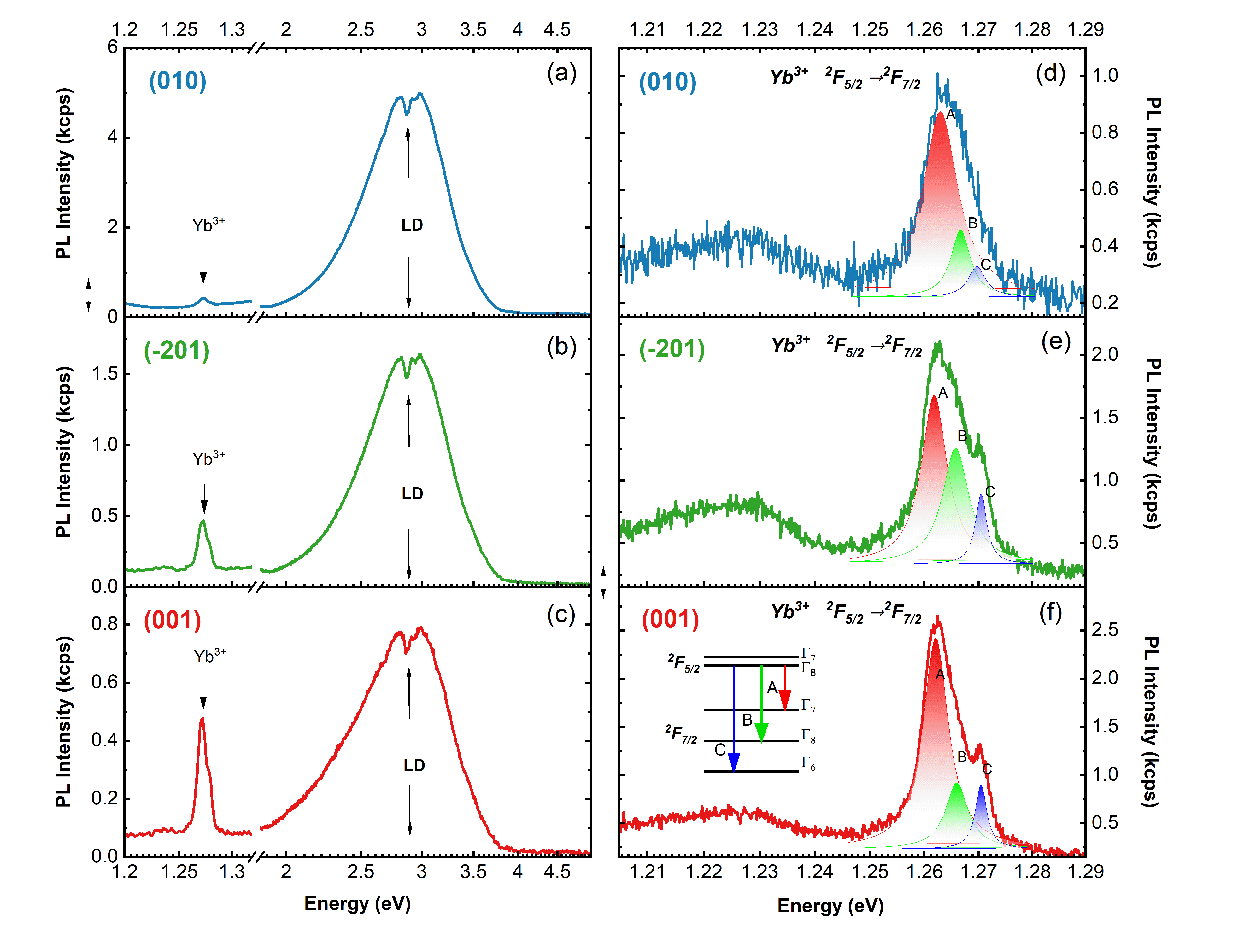}
    \caption{The PL spectra of (001), (010), and (-201)-oriented \bgao crystals, implanted with Yb ions with fluence of \fluence{15} and annealed in oxygen at 800\degree C for 10 min. The inset shows a diagram of radiative transitions between spin-orbital levels of 2F\textsubscript{5/2} → 2F\textsubscript{7/2 }terms of Yb\textsuperscript{3+} ions.}

    \label{fig:pl}
\end{figure}

Typical photoluminescence spectra collected for \bgao with Yb systems, under optical excitation at energy EL=5.82 eV above the band gap energy of \bgao, are shown in Figure \ref{fig:pl}. The spectra on the left represent the spectra collected in the wide spectral range from ultraviolet (UV) to near-infrared (NIR). The spectra on the right correspond to the NIR range, where the characteristic 4f inner shell transitions of the Yb ions in the 3+ oxidation state can be observed separately from the other characteristic emissions of \bgao\  \cite{prucnal2010blue}. 

As has already been reported, Yb ions remain optically inactive directly after implantation and annealing is necessary to get the optical response, and the most efficient luminescence of Yb-implanted \bgao is obtained after annealing in oxygen for 10 minutes at 800$\degree$ \cite{sarwar2024crystal}. Annealing under these conditions leads to the definite disappearance of the $\gamma$ phase and to the reduction of radiation-induced strain (see Figure \ref{fig:xrd_201}b). In the case of the highest fluence used i.e. \fluence{15} such annealing conditions lead to \bgao\ recrystallization in both the \ggao\ and the amorphous phases (see Figure SM3). 
However, due to the defects in the near-surface layer, \bgao\ recrystallization process is probably different (non-planar) than in the deeper part of the crystal, and the recovered \bgao\ layer might regrow misoriented with respect to the bulk crystal \cite{azarov2024optical}.
It could be the reason for the observed increase in the dechanneling level of the RBS-aligned spectra after annealing (see Figure SM4). However, the other factors such as dislocation migration, grain boundaries, and formation of RE-O and/or defect clusters cannot also be excluded. Consequently, further studies are essential for a comprehensive understanding of this phenomenon.

The PL spectra collected in a broad spectral range from UV to NIR (left part of Figure \ref{fig:pl}) do not reveal the characteristic near-band edge emission of \bgao\ around 4.8 eV, mostly due to the efficient interactions of free electrons with self-trapped holes and indirect energy bandgap \cite{yamaguchi2004first, guo2024gallium}. Instead, a broad PL band in the 3.8-2 eV range with a maximum of 2.8 eV associated with native local defects (LD band) and a narrow emission line of Yb\textsuperscript{3+} ions at 1.265 eV ($\lambda$=980 nm) are observed. In this case, the LD band is characterized by a short emission decay time of $\tau$\textsubscript{R}=7.8 ns. This value does not change over the spectral contour of this broadband. This fact allows us to state that this band is associated with a local defect of the same type and is formed by phonon modes of the emitter center. 
The origin of this center has not yet been established, but it could be associated with a divacancy or a vacancies complex, as well as a vacancy-impurity complex \cite{guo2024gallium, dong2017effects, sun2021oxygen, Varley2010_PL, BINET1998_PL}.

As can be seen in the right part of Figure \ref{fig:pl}, the Yb PL line is formed by the transition between the split-orbit levels of the excited 2F\textsubscript{5/2} and ground 2F\textsubscript{7/2} states of the Yb\textsuperscript{3+} ion (see the diagram of the transitions in Figure \ref{fig:pl}f). A weak PL band in the range of 1.2-1.24 eV is most likely associated with native local defects radiative transitions since this band is also observed in undoped samples of \bgao\ crystals.

For different orientations, strong differences in the PL intensity of Yb\textsuperscript{3+} ions are observed. At the same time, the energy and width of the spin-orbit components remain constant with the accuracy of the fit. The highest intensity of the Yb main peak is obtained for (001)-oriented \bgao. Recall that such oriented crystals exhibit the highest level of radiation-induced DIS-type defect, while the lowest level of this type of defect is observed for (010)-oriented crystals. Some reports \cite{ravadgar2013defects, esteves2023probing} indicate the important role of dislocations in enhancing luminescence in \bgao. Dislocations could be trapping centers for dopant ions, stabilizing them in the lattice at the dislocation core. Such location makes it easier to transfer energy from the crystal lattice to the ions, which enhances light emission from dopant \cite{zhou2010atomic}. Such a situation also may take place in this case, which would explain the high level of bending channel defects in (001) and (-201)-oriented crystals. However, other proof is necessary. Nevertheless, the correlation between LD and Yb emission is simple: the lower the defect excitation, the higher the Yb\textsuperscript{3+} luminescence.

\section{Conclusion}

The Yb-implanted and annealed \bgao\ crystals with (010), (001), and (-201) surface orientations were investigated with HRXRD, RBS/c, Raman spectroscopy and PL spectroscopy. The analyses of the results obtained by RBS/c were supplemented with MC simulations using McChasy.

The presented results clearly show the structural and optical anisotropy of Yb-implanted \bgao, with the different response of (010)-oriented crystals to the implantation process. The HRXRD 2$\theta$ scans for Yb implanted \bgao\ reveal that radiation-induced compressive strain develops for (010)-oriented crystals, while tensile strain for (001) and (-201)-oriented crystals was observed. 
The Raman measurements showed that the high-frequency phonon modes associated with the stretching and bending of Ga$_\mathrm{I}$O$_4$ tetrahedrons appear only for (010)-oriented crystals. 
Finally, the RBS/c studies, supported by McChasy simulations, demonstrated a significantly lower tendency to develop bending channel defects in the (010)-oriented crystals in comparison with the other tested crystal orientations. For (001) and (-201)-oriented Yb-implanted \bgao\ crystals the concentration of this type of defect is similar. The PL spectra obtained for (010)-oriented implanted crystal reveal the highest point-defect-related luminescence, which results in the low luminance efficiency from Yb\textsuperscript{3+}.  In contrast,  the high Yb\textsuperscript{3+}-related luminescence in (001) and (-201)-oriented crystals was observed. Thus, the results may suggest the important role of dislocations (or another bending channel defect) in the mechanism of Yb\textsuperscript{3+} luminescence in $\beta$-Ga$_2$O$_3$, which are probably trapping centers for dopant ions and/or defects leading to enhancing RE\textsuperscript{3+} emission. The mechanism is the subject of further research. This strong dependence of radiation-induced damage on the crystal orientation and different optical responses to these structural changes mean that (001) and (-201)-oriented \bgao\ can be recommended for optoelectronic applications that can be operated in harmful conditions e.g. high radiation environment and (010)-oriented \bgao\ for power electronics devices.

\medskip
\textbf{Acknowledgements} \par 
The research was carried out within the NCN project UMO-2022/45/B/ST5/02810. The experimental work was supported by the Helmholtz-Zentrum Dresden-Rossendorf (21002661-ST, 21002663-ST).

\medskip

\bibliographystyle{unsrt}
\bibliography{bibliography}

\includepdf[pages={-}]{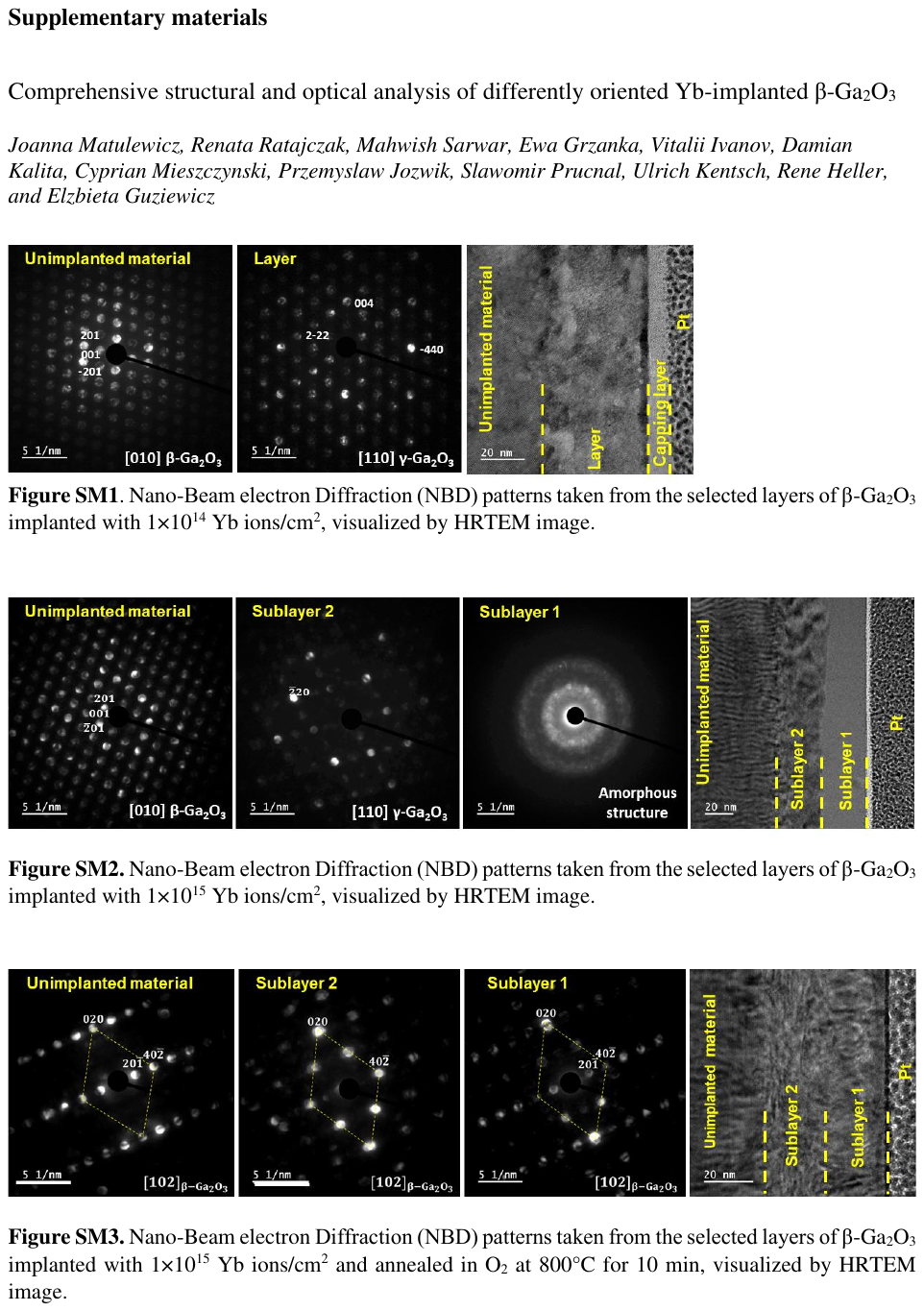}

\end{document}